\documentclass[a4paper,11pt]{article}
\usepackage{graphicx}
\usepackage{ifthen}
\usepackage{enumerate}
\usepackage{amsmath}
\usepackage{amssymb}
\usepackage{overpic}
\usepackage[mathscr]{eucal}
\usepackage[errorshow]{tracefnt}
\usepackage{amsmath} 
\usepackage{amssymb}
\usepackage{amscd} 
\usepackage{eucal}
\usepackage{hyperref}

\newcommand{\id}{\hbox{1\kern-.27em l}}

\begin{document}

{\hfill{\tt ULB-TH/10-24}}

\begin{center} {\bf \Large An M-theory solution from null roots in $E_{11}$}

\vspace{.9cm}

Laurent Houart, Axel
Kleinschmidt \\and Josef Lindman H\"ornlund

\footnotesize
\vspace{.9 cm}

{\em Service de Physique Th\'eorique et Math\'ematique,\\
Universit\'e Libre de Bruxelles \& International Solvay Institutes\\ Campus Plaine C.P. 231, Boulevard du
Triomphe, B-1050 Bruxelles, 
Belgium}

\vspace{.4 cm}

 {\tt
 lhouart, axel.kleinschmidt, jlindman@ulb.ac.be} 

\end{center}

\vspace {.7cm}

{\abstract

\noindent We find a purely gravitational classical solution of M-theory/eleven-dimensional supergravity which corresponds to a solution of the $E_{10}$ brane sigma-model involving a null root. This solution is not supersymmetric and is regularly embedded into $E_{11}$.}

\setcounter{equation}{0}

\section{Introduction}

Since the discovery of evidence for an hidden infinite symmetry in M-theory or eleven dimensional supergravity in terms of the Kac-Moody algebras $\mathfrak{e}_{10}$ or $\mathfrak{e}_{11}$ \cite{West:2001as, Damour:2002cu}, various attempts have been made to understand how, and if, these conjectured infinite symmetries are realised. Near a space-like singularity one can describe the billiard-like dynamics in terms of Weyl reflections in $\mathfrak{e}_{10}$ \cite{Damour:2000hv} (see \cite{Damour:2002et,Henneaux:2007ej} for reviews). Using the Kac-Moody/supergravity dictionary of \cite{Damour:2002cu} and \cite{Damour:2004zy}, various cosmological solutions of eleven dimensional gravity were derived in \cite{Kleinschmidt:2005gz}. Similar to the description of extremal black holes in terms of geodesics on a pseudo-Riemannian coset manifold \cite{Breitenlohner:1987dg, Bossard:2009at}, the Kac-Moody description was extended to describe supersymmetric brane solutions of eleven dimensional solutions in \cite{Englert:2003py,Englert:2004it,Englert:2007qb}, in terms of a mixed signature $E_{10}/K(E_{10})$ sigma-model. We recently extended this analysis to non-marginal supersymmetric bound states of the various branes in M-theory \cite{Houart:2009ya}. By embedding $\mathfrak{sl}(n, \mathbb{R})$ subalgebras in $\mathfrak{e}_{10}$ or $\mathfrak{e}_{11}$ in various ways, we re-derived several known supersymmetric brane bound states. Related earlier work in terms of $E_{11}$ group elements can be found in \cite{Cook:2009ri}.

After the considerations of \cite{Houart:2009ya}, we can now ask the question: What kind of supergravity solutions do different choices of subalgebras of $\mathfrak{e}_{10}$ or $\mathfrak{e}_{11}$ correspond to? In particular, which subalgebras give rise to supersymmetric solutions and which ones do not? The solutions studied earlier correspond to finite-dimensional subalgebras as they appear in the Cartan classification or to summands of $\mathfrak{gl}(1)$. This in particular included only algebra generators associated with {\em real} root generators or from the Cartan subalgebra.\footnote{In the case of $\mathfrak{e}_{10}$ and the so-called cosmological model, Borcherds algebras were also analysed~\cite{Kleinschmidt:2005gz}.}
By contrast, we will here  consider a non-simple Lie algebra $\mathfrak{g} \subset \mathfrak{e}_{11}$ that involves generators associated to a {\em null} root of $\mathfrak{e}_{11}$.\footnote{The role of purely imaginary (time-like) root generators is largely unknown, see, however, the analysis of~\cite{Brown:2004jb}.} Using the brane sigma-model we derive from this extended Heisenberg algebra a static space-time, that turns out to break supersymmetry completely. This example illustrates the fact that different types of subalgebras of $\mathfrak{e}_{10}$ or $\mathfrak{e}_{11}$ correspond to a wide range of different solutions in M-theory.  

We will, in the following, assume some familiarity with the brane sigma-model. For an introduction on how subalgebras in $\mathfrak{e}_{10}$ or $\mathfrak{e}_{11}$ map to supergravity solutions via the Kac-Moody/supergravity dictionary, we refer to the more extensive analysis in \cite{Houart:2009ya}. One advantage of analysing supergravity solutions from the algebraic perspective is that the sigma-models that are conjectured to be dual to the gravitational theory are formally completely integrable and hence in principle allow for the construction of solutions with arbitrary conserved Noether charges.

\setcounter{equation}{0}
\section{A null root solution}

The method used here and in \cite{Kleinschmidt:2005gz, Houart:2009ya} goes as follows. One first picks a suitable algebra $\mathfrak{g}\subset\mathfrak{e}_{11}$ (not necessarily simple)  and determines the null geodesics on $G/K$ where $G$ is the Lie group associated with $\mathfrak{g}$ and $K$ a maximal subgroup whose real form depends on whether one is interested in cosmological or brane type solutions in the end. Then one finds a realisation of this algebra in terms of the generators at lower levels in the level decomposition of $\mathfrak{e}_{11}$. Via the dictionary of for example \cite{Damour:2004zy} one then derives the corresponding supergravity solution and analyses its properties. In this short note we choose $\mathfrak{g}$ to be a four-dimensional nilpotent Lie algebra, containing a Heisenberg subalgebra with a null root.

\subsection{A nilpotent Lie algebra}

Recall the Cartan-Weyl basis of a general Lie algebra,
\begin{equation}
\label{eqn:nilalgebra}
[h_{\alpha}, e_{\pm \beta}] = \pm \beta(h_{\alpha}) e_{\pm \beta}, \quad [e_{\alpha}, e_{-\beta}] = \delta_{\alpha, \beta} h_{\alpha}. 
\end{equation}
The Killing form $\kappa$ restricted to the Cartan subalgebra is in this basis given by 
\begin{equation}
\label{eqn:killing}
(h_{\alpha},h_{\beta}):=\kappa(h_{\alpha}, h_{\beta}) = \alpha(h_{\beta}) .
\end{equation}
We call a root $\alpha$ null (or light-like imaginary) if $\alpha(h_{\alpha}) = 0$, i.e. if it has zero norm under the Killing form (\ref{eqn:killing}). This implies that $h_{\alpha}$ commutes with the positive and negative step operators $e_{\pm\alpha}$ corresponding to the same root $\alpha$. If $\alpha(h_{\alpha}) < 0$ we call $\alpha$ purely imaginary (or time-like imaginary). The existence of purely imaginary and null roots in Kac-Moody algebras is one of the things that distinguishes them from finite simple Lie algebras \cite{Kac:book}. (The root spaces associated with null or imaginary roots are in general degenerate.)

Let us now consider an algebra, that we denote $\mathfrak{g}$, generated by the four elements $h, e, f$ and $\Lambda$ subject to the commutation relations
\begin{equation}
\label{eqn:commutationrelations}
[h, e] = 0, \quad [h, f]=0,  \quad [\Lambda, e] = \mu e, \quad [h, \Lambda] = 0, \quad [e,f] = h, \quad [\Lambda, f] = -\mu f .
\end{equation}
Here we have not specified the normalisation of $\Lambda$ and this shows up in the arbitrary parameter $\mu$. This algebra is nilpotent, as 
\begin{equation}
[\mathfrak{g},[\mathfrak{g},[\mathfrak{g}, \mathfrak{g}]]] = 0,
\end{equation}
and $h$ is a center element, being in the centralizer of $\Lambda, e$ and $f$. In $\mathfrak{e}_{11}$ we may realise this algebra by choosing for example\footnote{This algebra becomes a subalgebra of $\mathfrak{e}_{10}$ if we truncate away the generator ${K^1}_1$ corresponding to the node in $\mathfrak{e}_{11}$ but not in $\mathfrak{e}_{10}$. As the resulting space-time solution is identical, we can choose to consider either the $E_{11}$ or the $E_{10}$ sigma-model.}
\begin{eqnarray}
e &=& R^{3|4567891011}, \nonumber \\
\label{eqn:aine10}
f &=& R_{3|4567891011},  \\
h &=& -{K^1}_1-{K^2}_2, \nonumber  \\
\Lambda &=& \mu \left( {K^1}_1+\frac{1}{9}\sum_{a=3}^{11} {K^a}_a \right) \nonumber.
\end{eqnarray}
Here we follow the conventions of \cite{Damour:2004zy} and \cite{Englert:2004ph}. Recall that in the level decomposition of $\mathfrak{e}_{11}$ under the gravity $\mathfrak{gl}(11, \mathbb{R})$ subalgebra, ${K^{a}}_b$ span the zeroth level, and the generators $R^{a_0|a_1...a_8}$ span level three (see e.g. \cite{Henneaux:2007ej} for more details). We choose the following bilinear form on $\mathfrak{g}$, 
\begin{equation}
\label{eqn:bilinearform}
(e,f) = 1, \quad (\Lambda, \Lambda) = -\frac{8}{9}\mu^2, \quad (h,\Lambda) = \mu,
\end{equation}
which is consistent with the standard bilinear form of $\mathfrak{e}_{11}$ according to the embedding (\ref{eqn:aine10}). The bilinear form vanishes on all other combinations and in particular $(h,h) = 0$. Hence, $h$ is a Cartan generator corresponding to a null root in $\mathfrak{e}_{11}$, as can also be seen by the commutation relations (\ref{eqn:commutationrelations}). Note that the bilinear form is non-degenerate, due to the non-zero `angle' $\mu$ between $h$ and $\Lambda$. Define furthermore an involution $\tau$ on $\mathfrak{g}$ by
\begin{equation}
\label{eqn:temporalinvolution}
\tau(h) = -h, \quad \tau(\Lambda) = -\Lambda ,\quad \tau(e) = f,
\end{equation}
such that the fixed point set under $\tau$ is the subalgebra
\begin{equation}
\mathfrak{k} = \mathbb{R}(e+f), 
\end{equation}
and 
\begin{equation}
\mathfrak{p} = {\rm span}_{\mathbb{R}} ( h, \Lambda, e-f),
\end{equation}
is the set of elements $X \in \mathfrak{g}$ that obey $\tau(X) = -X$. If we think of $\mathfrak{g} \subset \mathfrak{e}_{11}$, according to (\ref{eqn:aine10}), $\tau$ is the restriction of the `temporal' involution of \cite{Englert:2003py}, by letting for example the eleventh direction be time-like. 

\subsection{Solving the sigma-model equations of motion}

Let us now consider a sigma-model generated by a map from $\mathbb{R}$, parametrized by the co-ordinate $\xi$, into the coset manifold $G/K$, where $G$ is the Lie group with algebra $\mathfrak{g}$ and $K$ is the subgroup of $G$, generated by the subalgebra $\mathfrak{k}$. From the bilinear form (\ref{eqn:bilinearform}) we see that $K = \mathrm{SO}(1,1)$, as $(e+f,e+f) = 2$, indicating that $e+f$ is a non-compact generator. We let (locally on $G$)
\begin{equation}
\mathcal{V} = \exp (\phi h + q \Lambda)\exp(Ae),
\end{equation}
be the coset map and $\phi$, $q$ and $A$ depend on the parameter $\xi$. We denote by $\mathcal{P}$ and $\mathcal{Q}$ the usual projections to $\mathfrak{p}$ and $\mathfrak{k}$ respectively, of the pullback $\mathcal{V}^*(\omega_G)$, where $\omega_G$ is the Maurer-Cartan form on $G$. More concretely
\begin{equation}
\mathcal{P} = \frac{1}{2}(\id - \tau)(\partial_{\xi} \mathcal{V} \mathcal{V}^{-1}), \quad \mathcal{Q} = \frac{1}{2}(\id + \tau)(\partial_{\xi} \mathcal{V} \mathcal{V}^{-1}), 
\end{equation}
and we find
\begin{equation}
\label{eqn:cosetvelocity}
\mathcal{P} = \partial_{\xi} \phi h + \partial_{\xi} q \Lambda + e^{\mu q} \partial_{\xi} A \frac{e-f}{2},
\end{equation}
and 
\begin{equation}
\mathcal{Q} =  e^{\mu q} \partial_{\xi} A \frac{e+f}{2} .
\end{equation}
Demanding that $\mathcal{V}$ describes a geodesic on $G/K$ amounts to
\begin{equation}
\partial_{\xi} \mathcal{P} - [\mathcal{Q}, \mathcal{P}]=0,
\end{equation}
and in terms of $q, \phi$ and $A$ this gives us the three equations
\begin{eqnarray}
\partial_{\xi}^2 q& =& 0, \nonumber \\
\label{eqn:diffequations}
\partial_{\xi}^2 \phi + \frac{1}{2} e^{2\mu q} (\partial_{\xi} A)^2& = &0 ,\\
\partial_{\xi}(e^{2\mu q} \partial_{\xi} A) &=& 0. \nonumber
\end{eqnarray} 
These equations are quite straightforwardly integrated and the solution is given by
\begin{eqnarray}
q &=& c_1 \xi + c_2 \nonumber, \\
\label{eqn:solution}
A &=& c_3e^{-2\mu (c_1 \xi + c_2)}+c_6 ,\\
\phi &=& - \frac{c_3^2}{2}e^{-2\mu( c_1 \xi + c_2)}+c_4 \xi + c_5, \nonumber
\end{eqnarray}
where $c_1,\ldots, c_6$ are six integration constants. The lapse constraint, derived from reparametrization invariance on the world-line spanned by $\xi$ (and ensuring that $\mathcal{V}$ traces out a null geodesic on $G/K$) is
\begin{equation}
(\mathcal{P}, \mathcal{P}) = 0,
\end{equation}
and this condition becomes, using (\ref{eqn:cosetvelocity}) and the bilinear form (\ref{eqn:bilinearform}), 
\begin{equation}
\label{eqn:lapseconstraint}
-\frac{8\mu^2}{9}(\partial_{\xi} q)^2 +2\mu \partial_{\xi} q\partial_{\xi} \phi - \frac{1}{2} e^{2\mu q}(\partial_{\xi} A)^2 = 0 .
\end{equation}
We solve (\ref{eqn:lapseconstraint}) by setting $c_4 = \frac{4}{9}\mu c_1$.

Note that the equations (\ref{eqn:diffequations}) and (\ref{eqn:lapseconstraint}) incorporate a couple of symmetries, corresponding to the four generators of the Lie algebra $\mathfrak{g}$. We have two shift-symmetries, $\phi \rightarrow \phi+a$ and $A \rightarrow A+b$ as $\phi$ and $A$ only show up with their differentials $\partial_{\xi}\phi$ and $\partial_{\xi} A$ in (\ref{eqn:diffequations}) and (\ref{eqn:lapseconstraint}). These two symmetries shift the $c_6$ and $c_5$ integration constants. Associated to the generator $\Lambda$ is the symmetry, 
\begin{eqnarray}
q &\rightarrow &q+ a ,\nonumber\\
A &\rightarrow &e^{  -\mu a}A \nonumber,
\end{eqnarray}
which acts on the integration constants as $c_2 \rightarrow c_2 + a$, $c_3\rightarrow e^{\mu a} c_3$ and $c_6 \rightarrow e^{-\mu a} c_6$. We also have a non-linear symmetry, corresponding to the non-compact subgroup $K = \left\{ e^{\alpha (e+f)}\,:\, \alpha\in\mathbb{R}\right\}$ acting as
\begin{eqnarray}
\phi &\rightarrow& \phi - \frac{1}{2}\alpha^2 e^{-2\mu q} + \alpha \partial_{\xi} A, \\
A &\rightarrow &A + \alpha e^{-2\mu q} .
\end{eqnarray}
This symmetry acts on the $c_3$ constant as $c_3 \rightarrow c_3+\alpha$ and hence is equivalent to switching on the level three generator. Note that the four-dimensional symmetry group $G$ does not act transitively on the five-dimensional space of solutions, since there is no symmetry that acts on the integration constant $c_1$. In other words, on the space of solutions to the sigma-model, the parameter $c_1$ parametrises the orbits of $G$.

\subsection*{Dictionary}

We can embed the above solution in the $E_{10}$ brane sigma model, and use the standard dictionary of \cite{Damour:2002cu,Damour:2004zy} (see also \cite{Damour:2006xu}) to derive the corresponding space-time solution. We fix time $t$ to be $x_{11}$ and the parameter $\xi$ in the sigma-model to map to the $x_1$ coordinate. Let ${e_m}^a$ be the (ten-dimensional) vielbein in the directions orthogonal to $\xi$, and assume that ${e_m}^a$ is diagonal. Here, $m$ denotes a curved index; we will also use the notation where we put a tilde over a curved index, especially for a specific value. This then gives
\begin{equation}
{e_{\tilde{2}}}^2 = e^{\phi},
\end{equation}
from the dictionary for $\phi$. Furthermore, the dictionary for the field $q$ gives 
\begin{equation}
{e_{\tilde{b}}}^b= e^{-\frac{\mu}{9}q+ \chi_b},
\end{equation}
where $b = 3,...,11$ and the $\chi_b$'s are functions, not depending on $\xi$. From the dictionary for the $A$ field we will see below that the $\chi_b$'s are necessarily non-zero. Let $\Omega_{abc}$ be the anholonomy derived from the vielbein ${e_m}^a$. The dictionary of \cite{Damour:2004zy} demands that the trace of the last two indices in the anholonomy vanishes, i.e.
\begin{equation}
\label{eqn:trace}
0 = \Omega_a \equiv   \sum_{b=2}^{11} \Omega_{abb} .
\end{equation}
If we furthermore assume that the $\chi_b$ are functions of $x_2$ only, the trace condition, taking $a=2$ in (\ref{eqn:trace}), implies that
\begin{equation}
\label{eqn:zerotrace}
\sum_{b=3}^{11}\partial_{\tilde{2}} \chi_b = 0 .
\end{equation}
Finally, the dictionary for the $A$-field gives
\begin{equation}
\label{eqn:Adictionary}
2 e \Omega_{233} = e^{-2\mu q} \partial_{\xi} A,
\end{equation}
where $\Omega_{233}$ is a component of the anholonomy $\Omega_{abc}$ derived from the vielbein ${e_m}^a$, and $e = \det {e_m}^a$. If we choose 
\begin{equation}
\sum_{b=3}^{11} \chi_b = 0, 
\end{equation}
the $\chi_b$ dependence disappears from the determinant of the vielbein, and we find from (\ref{eqn:Adictionary}) that
\begin{equation}
\chi_3 = \mu c_1 c_3 x_2,
\end{equation}
where $x_2$ is the co-ordinate in the $2$-direction. We can solve the trace-condition (\ref{eqn:zerotrace}) by choosing $\chi_4 = -\chi_3$ and $\chi_c = 0$ for $c\geq 5$. Due to the fact that the level 3 generator in the $\mathfrak{e}_{11}$ embedding has no leg in the $x_1$-direction, we have hence found from equation (\ref{eqn:Adictionary}) that we need a dependence on $x_2$ in the metric, in addition to the dependence of $\xi$. The last component of the vielbein is now given by the dictionary for the lapse function in the sigma-model, and becomes
\begin{equation}
\label{eqn:lastdictionary}
{e_{\xi}}^{1} = e = e^{\phi -\mu q}.
\end{equation}

\subsection{Space-time solution}

To summarize, from the algebra $\mathfrak{g}$ given by (\ref{eqn:nilalgebra}) we have found, using the dictionary of \cite{Damour:2004zy}, the space-time metric
\begin{eqnarray}
\mathrm{d}s_{11}^2 &= &e^{2\phi-2\mu q}\mathrm{d}\xi^2 +e^{2 \phi}\mathrm{d}x_2^2 + \sum_{b=5}^{10} e^{-\frac{2\mu}{9}q} \mathrm{d}x_b^2 \nonumber \\
\label{eqn:spacetime}
& & + e^{-\frac{2\mu}{9}q+2 \mu c_1 c_3 x_2} \mathrm{d}x_3^2+e^{-\frac{2\mu}{9}q-2 \mu c_1 c_3 x_2} \mathrm{d}x_4^2  -e^{-\frac{2\mu}{9}q}\mathrm{d}t^2.
\end{eqnarray}
This is a Ricci-flat metric, and is therefore a solution of the classical equations of motion in eleven dimensional supergravity. There are no fluxes. We note that this solution is very similar to polarised Gowdy cosmologies, although stationary. Some Gowdy-type cosmologies were found using Heisenberg-subalgebras in \cite{Kleinschmidt:2005bq} in the cosmological Kac-Moody sigma-model for pure gravity. Below we analyse the amount of supersymmetry preserved by the space-time defined by (\ref{eqn:spacetime}).

By putting the level three field $A$ to zero, i.e. $c_3=0$ in (\ref{eqn:solution}), the solution reduces to the stationary version of the Kasner-solution \cite{Landau:1987}, which has been discussed in the context of over-extended Kac-Moody algebras in \cite{Englert:2003zs}. The stationary Kasner solution is known not to be supersymmetric. Using the $K$-symmetry discussed above, we can generate the full solution with the level three generator switched on, from the stationary Kasner and as $K$-transformations are expected to preserve supersymmetry we can deduce that the full solution breaks supersymmetry as well. We confirm this below.

\subsubsection*{Brane embedding}

As discussed in \cite{Englert:2004ph}, a solution of the $E_{10}$ brane sigma-model can be embedded in an $E_{11}$ sigma-model if the relation
\begin{equation}
\label{eqn:embedding}
g_{\xi \xi}= e^2,
\end{equation}
is satisfied, where $e$ is the determinant of the vielbein in the directions orthogonal to $\xi$. The relation (\ref{eqn:embedding}) is similar to the brane extremality condition of \cite{Argurio:1997gt}, and is automatically satisfied by (\ref{eqn:lastdictionary}).

\subsection*{Checking supersymmetry}
\label{sec:checkingsusy}

It is now straightforward to confirm that the solution (\ref{eqn:spacetime}) derived from the nilpotent Lie algebra $\mathfrak{g}$ breaks supersymmetry completely. For clarity we do the analysis quite explicitly. The equation for a vanishing variation of the gravitino in eleven dimensional supergravity is
\begin{equation}
\label{eqn:gravitinovariation}
(\partial_M + \frac{1}{4}{\omega_M}^{AB}\Gamma_{AB})\epsilon = 0, 
\end{equation}
if all fluxes are set to zero, as in the above solution. Here ${\omega_M}^{AB}$ is the spin-connection, $\epsilon$ a Majorana spinor (32 independent real components), and $\Gamma_{A}$ the generators of the Clifford algebra in 10+1 dimensions. The components of the spin connection, calculated from (\ref{eqn:spacetime}), become (in flat indices)
\begin{eqnarray}
\omega_{bb1}& =& -\frac{\mu}{9}\eta_{bb}e^{-\phi+\mu q}\partial_{\xi} q \quad(\text{no sum})\nonumber ,\\
\omega_{221} &= &e^{-\phi+\mu q} \partial_{\xi}\phi ,\\
\omega_{332} & = & e^{-\phi}  \mu c_1 c_3  \nonumber ,\\
\omega_{442} & = & -e^{-\phi}  \mu c_1 c_3 \nonumber  ,
\end{eqnarray}
and $b = 2,...,11$. The first three non-zero parts correspond to the three fields we have in the sigma-model, and $\omega_{442}$ was forced to be non-zero by the trace-constraint. All of the other components vanish. Hence, if we take $M = \xi=1$ in (\ref{eqn:gravitinovariation}), we see that ${\omega_{\xi}}^{AB} =0 $  and (\ref{eqn:gravitinovariation}) become
\begin{equation}
\label{eqn:epsilon}
\partial_{\xi} \epsilon = 0 .
\end{equation}
We continue with the case $M = \tilde{b}$, $b>4$. Here $\partial_{\tilde{b}} \epsilon = 0$ as the metric is independent of $x_b$. Hence
\begin{equation}
\frac{1}{4} {\omega_b}^{b1} \Gamma_{b1} \epsilon = 0 \quad (\text{no sum}),
\end{equation}
implying that $\Gamma_{b1} \epsilon = 0$ for all $b = 5,...,11$. Furthermore for $M = 3$ we have
\begin{equation}
\frac{1}{4} ({\omega_3}^{31} \Gamma_{31}+{\omega_3}^{32} \Gamma_{32}) \epsilon =0
\end{equation}
implying in particular that $\Gamma_{31} \epsilon = 0$ and $\Gamma_{32} \epsilon = 0$ . We clearly find a similar result for $M=4$ implying that $\epsilon = 0$.

\section{Conclusions and discussion}

We have seen in this short note that non-simple subalgebras of $\mathfrak{e}_{11}$ with imaginary roots correspond to solutions qualitatively very different from the algebras discussed in \cite{Houart:2009ya} that consisted only of real roots. In particular the solution discussed here is instead quite similar to cosmological solutions one can derive from `null algebras' in the cosmological $E_{10}$ sigma model. That the difference between the cosmological and the brane sigma model is quite minor in this case is evident from the lapse equation (\ref{eqn:lapseconstraint}). The lapse constraint for the algebra $\mathfrak{g}$ only restricts the integration constants of the solution, and not the functional dependence of the fields, contrary to the case in \cite{Houart:2009ya}.

To extend the results here, it would be interesting to consider null algebras with also level one and level two generators switched on. These generators turn on charges in space-time and it would be interesting to study the corresponding solutions, in order to see whether they are supersymmetric.

\subsection*{Acknowledgements}

\noindent
We would like to thank Riccardo Argurio for enlightening discussions. LH is a Senior Research Associate and AK is a Research Associate of the Fonds de la Recherche Scientifique-FNRS, Belgium. 
This work has been supported in part by IISN-Belgium (conventions 4.4511.06, 4.4505.86 
and 4.4514.08) and by the Belgian Federal Science Policy Office through the Interuniversity 
Attraction Pole P6/11. 

 \addcontentsline{toc}{section}{References}
\bibliographystyle{utphys}
\bibliography{Refdata}
\end{document}